\documentclass[aps,prl,twocolumn,groupedaddress,amsmath,amssymb]{revtex4-1}
\newcommand\thintilde{{\lower.74ex\hbox{\mathtt{\char`\~}}}}
\usepackage{graphics}
\usepackage{color}
\usepackage{graphicx}
\usepackage{natbib}
\newcommand{\un}[1]{\ensuremath{\, \mathrm{#1}}}
\newcommand{\RNum}[1]{\uppercase\expandafter{\romannumeral #1\relax}}
\begin{document}

\title{Transient scaling and resurgence of chimera states in networks of \\Boolean phase oscillators}

\author{David P. Rosin,$^{1,2}$ Damien Rontani,$^{1,3}$ Nicholas D. Haynes,$^1$ Eckehard Sch\"oll,$^2$ and Daniel J. Gauthier$^1$}

\affiliation{$^1$Department of Physics, Duke University, 120 Science Drive, Durham NC~27708, USA\\
$^2$Institut f\"ur Theoretische Physik, Technische Universit\"at Berlin, Hardenbergstr.~36, Berlin D-10623, Germany\\
$^3$Sup\'{e}lec, OPTEL research group and LMOPS EA-4423, 2 Rue Edouard Belin, Metz F-57070, France}
\thanks{}
\date{\today}

\begin{abstract}
We study networks of non-locally coupled electronic oscillators that can be described approximately by a Kuramoto-like model. The experimental networks show long complex transients from random initial conditions on the route to network synchronization. The transients display complex behaviors, including resurgence of chimera states, which are network dynamics where order and disorder coexists. The spatial domain of the chimera state moves around the network and alternates with desynchronized dynamics. The fast timescale of our oscillators (on the order of $100\un{ns}$) allows us to study the scaling of the transient time of large networks of more than a hundred nodes, which has not yet been confirmed previously in an experiment and could potentially be important in many natural networks. We find that the average transient time increases exponentially with the network size and can be modeled as a Poisson process in experiment and simulation. This exponential scaling is a result of a synchronization rate that follows a power law of the phase-space volume.
\end{abstract}


% 05.45.Xt : Synchronization; coupled oscillators
% 64.60.aq : Networks
% 84.30.Ng : electronics: Oscillators, pulse generators, and function generators
% 84.30.Sk : electronics: Pulse and digital circuits
% 89.75.Da : Scaling Phenomena in complex systems
% 05.45.-a : Nonlinear Dynamics

\maketitle

As discovered recently, the dynamical state of networks can show a surprising behavior called chimeras, where network nodes split into coexisting domains of entirely different dynamics, such as synchronized and unsynchronized dynamics~\cite{KUR02a,ABR04,ABR08,OME10a,OME11,ZAK14,OME13}. Chimera states have possible applications to brain activity patterns, cardiac fibrillation, and social systems \cite{PAN14}. Recently, chimera states have been identified theoretically as long chaotic transients towards synchrony for finite-size networks, scaling exponentially with the system size \cite{WOL11}. Such an exponential scaling of the transient with the system size is called a supertransient in extended systems, but is not commonly known to appear in networks. 

Predicting the timescale for the transition to synchronization is crucial for technological applications, such as cascading failure in power grids and turbulent flows in pipes \cite{SIM08,TEL08}. Furthermore, it is of great importance for biological systems, such as ecological and neural systems \cite{HAS94,RAB08}. Transient scaling is especially important in networks because network structures dominate many natural and engineered systems \cite{NEW10}, but has not yet been shown experimentally. For example, for networks displaying chimera states, slow characteristic timescales have previously prevented the measurement of transient scaling \cite{HAG12,TIN12,MAR13,LAR13,WIC13,SCH14a}. Even in theoretical studies, this scaling could only be verified in small networks of less than 45 nodes because larger networks require prohibitively long computation times \cite{WOL11,MOT13}. 

Here, we study the transient behavior of networks showing chimera dynamics in an experimental network of Boolean phase oscillators realized with electronic logic circuits. Because these nodes operate on a timescale of ${\sim}100\un{ns}$, we can study the scaling of the transient in large networks of more than a hundred nodes. The transient includes chimera states for about $14\%$ of the time for $N=128$ and ends in a nearly synchronized state. We find that the transient time follows a Poisson process with an average transient time that increases exponentially with the network size, which is a result of the synchronization rate that follows a power law of the phase-space volume.

The oscillatory network nodes are realized with unclocked logic circuits and directly-wired links on microelectronic chips, realizing an autonomous Boolean network (ABN). Besides their application as engineered systems for random number generation and neuromorphic computation \cite{ROS13,ROS13a,ROS13b}, ABNs are also a common model for genetic circuits \cite{KAU69,GHI85,GLA98}. The ABN studied here is a variant of all-digital phase-locked loops, which are widely used for frequency synthesis \cite{AL-06,BES03}.

\begin{figure}[tb]
\begin{center}
\includegraphics{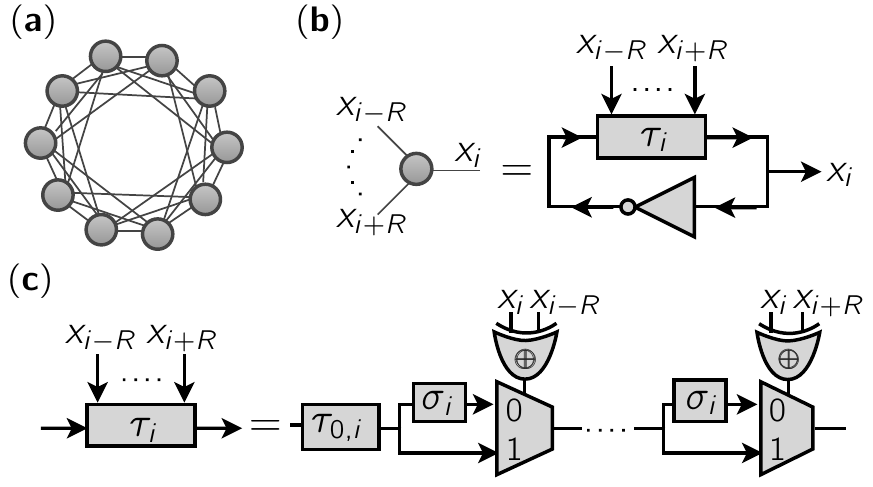}
\end{center}
\caption{\label{fig:bpo_node_setup}
(a) Illustration of a ring network with $N=10$ nodes and coupling range $R=3$. (b) Illustration of the Boolean phase oscillator (a node in the network) with state variable $x_i$. (c) The state-dependent delay for the coupling mechanism consisting of a constant delay $\tau_{0,i}$ built with 30 cascaded copier logic gates and $2R$ variable delay elements. $\sigma_i$, trapezoids, and $\oplus$-signs denote delay lines, Boolean switches (multiplexers), and XOR gates, respectively. }
\end{figure}

We study networks of $N$ coupled Boolean phase oscillators as shown schematically in Fig.~\ref{fig:bpo_node_setup}(a), where oscillator $i\in\left\{1,2,...,N\right\}$ is non-locally coupled to multiple other oscillators $j$ forming a network. The oscillators consists of an inverter gate with delayed feedback as shown in Fig.~\ref{fig:bpo_node_setup}(b). For constant delay $\tau_i$, this setup is known as a ring oscillator with a frequency given by \cite{KAT98}
\begin{equation}\label{eq:ring_osc}
 f_i = \frac{1}{2\tau_i},
\end{equation}
where the factor two accounts for inverted delayed feedback (one period includes two inversions). We extend the oscillator to allow for an adjustable frequency by making the delay $\tau_i$ state-dependent so that $\tau_i$ and $f_i$ change in response to the coupling signals. The coupling signals are generated by measuring the phase difference between the local oscillator and its neighbors, as introduced in Ref.~\cite{ROS14} for two oscillators.

%%%%%%%%%%%%%%%%%%%%%%%%%%%%%%%%%%%%%%%%%%%%%%%%%%%%%%%%%%%%%%%%
%%%%%%%%%%%%%%%%%%%%%%%%%%%%%%%%%%%%%%%%%%%%%%%%%%%%%%%%%%%%%%%%%%%

The state-dependent delay $\tau_i$ of an oscillator is built from unclocked logic gates as shown in Fig.~\ref{fig:bpo_node_setup}(c). It includes a constant delay $\tau_{0,i}$ \cite{ROS14}, and a variable delay realized with a combination of XOR logic gates, Boolean switches, and short constant delay lines $\sigma_i$. The XOR logic gates generate a signal $x_{i} \oplus x_j$ that approximates the phase difference between the $i$th oscillator ($x_i$) and its $j$th neighbor oscillator ($x_j$). This signal activates one of two paths in the setup of which one has an additional constant delay. When all phase differences are zero ($x_i \oplus x_j=0$ for all $j$), then the maximum feedback delay is selected with $\tau_i = \tau_{0,i}+2R\sigma_i$. When, on the other hand, a phase difference is detected ($x_i \oplus x_j=1$), the delay decreases by $\sigma_i$. This behavior can be expressed with state-dependent delay
\begin{equation}\label{eq:state_dependent_delay}
\tau_i=\tau_{0,i}+\sigma_i\sum_{\substack{j=i-R\\j\neq i}}^{i+R}{(1-x_i\oplus x_{j})},
\end{equation}
which is inserted in Eq.~\eqref{eq:ring_osc} to approximate the frequency adjustment, leading to a coupling mechanism of oscillators and hence the possibility of synchronization~\cite{ROS14}.

As detailed in the supplementary material \cite{supplement}, combining Eqs.~\eqref{eq:ring_osc} and \eqref{eq:state_dependent_delay} leads to an approximate phase model for the Boolean phase oscillators
\begin{equation}\label{eq:phase_model2}
\dot{\phi}_i = \omega_{0,i} + \tilde{\sigma}_i\sum_{j=i-R}^{i+R}\left|\Theta\left[\sin(\phi_j)\right] -\Theta\left[\sin(\phi_i+\alpha_{ij})\right]\right|,
\end{equation} 
with free-running frequencies $\omega_{0,i}$, coupling strengths $\tilde{\sigma}_i$, phase lag parameter $\alpha_{ij}$ that results from transmission delays, and Heaviside function $\Theta$.
The oscillators are non-locally coupled in a ring network with a coupling range $R$ as shown schematically in Fig.~\ref{fig:bpo_node_setup}(a). This configuration has been used previously to observe chimera states with the Kuramoto model, which is similar to Eq.~\eqref{eq:phase_model2} \cite{KUR02a,ABR04,KUR84,WOL11,OME10a,OME11,OME13,ZAK14}. 

The experimental oscillators have an intrinsic frequency heterogeneity of $\left|\sigma_f\right|/\bar{f} = 0.3\%$ with average frequency $\bar{f}=9.14\un{MHz}$ and standard deviation $\sigma_f=0.03\un{MHz}$ \cite{supplement}. In the model, we assume identical oscillators ($\omega_{0,i}=\omega_{0}$) and homogeneous coupling ($\tilde{\sigma}_i=\tilde{\sigma}$ and $\alpha_{ij}=\alpha$). 

\begin{figure}[tb]
\begin{center}
\includegraphics[width=\linewidth]{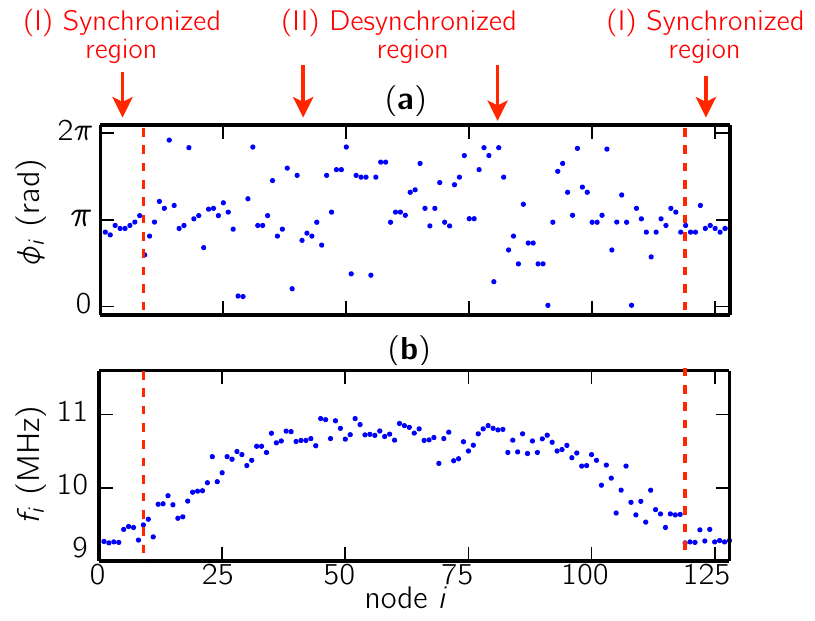}
\end{center}
\caption{\label{fig:exp_dynamics}
(Color online) Dynamics measured from coupled Boolean phase oscillators with $N=128$, $R=30$, $\omega_0 = 2\pi(9.3\pm0.03)\un{MHz}$, $\tilde{\sigma} = 2\pi(0.089 \pm 0.003)\un{MHz}$. 
(a) Snapshot at $t\approx 304\un{s}$, (b) frequency profile $f_i=\langle\dot{\phi_i}\rangle/(2\pi)$. The network is initialized by deactivating the coupling, resulting in randomized initial phases, followed by activating the coupling. $i$ is shifted by a constant to center the unsynchronized domain \cite{supplement}.}
\end{figure}

We first describe a part of the network dynamics in Fig.~\ref{fig:exp_dynamics}(a), showing a snapshot of the phase of oscillators in a chimera state. The oscillators outside (inside) the dotted lines, marked region \RNum{1} (region \RNum{2}), have equal (different) phases within our measurement precision of $\Delta\phi=\pm0.25\un{rad}$ and hence are considered phase synchronized (desynchronized). Therefore, the oscillators in region \RNum{1} stay synchronized, whereas those in region \RNum{2} drift apart because they have different frequencies. These frequencies are shown in Fig.~\ref{fig:exp_dynamics}(b) and are measured over a time period of $6\un{\mu s}$, which represents approximately 60 oscillation periods with precision of $\pm0.2\un{MHz}$. The oscillators in region \RNum{2} show the characteristic spectral feature of chimera states \cite{KUR02a,ABR04}. 

%%%%%%%%%%
%%%%%%%%%%

\begin{figure}[tb]
\begin{center}
\includegraphics{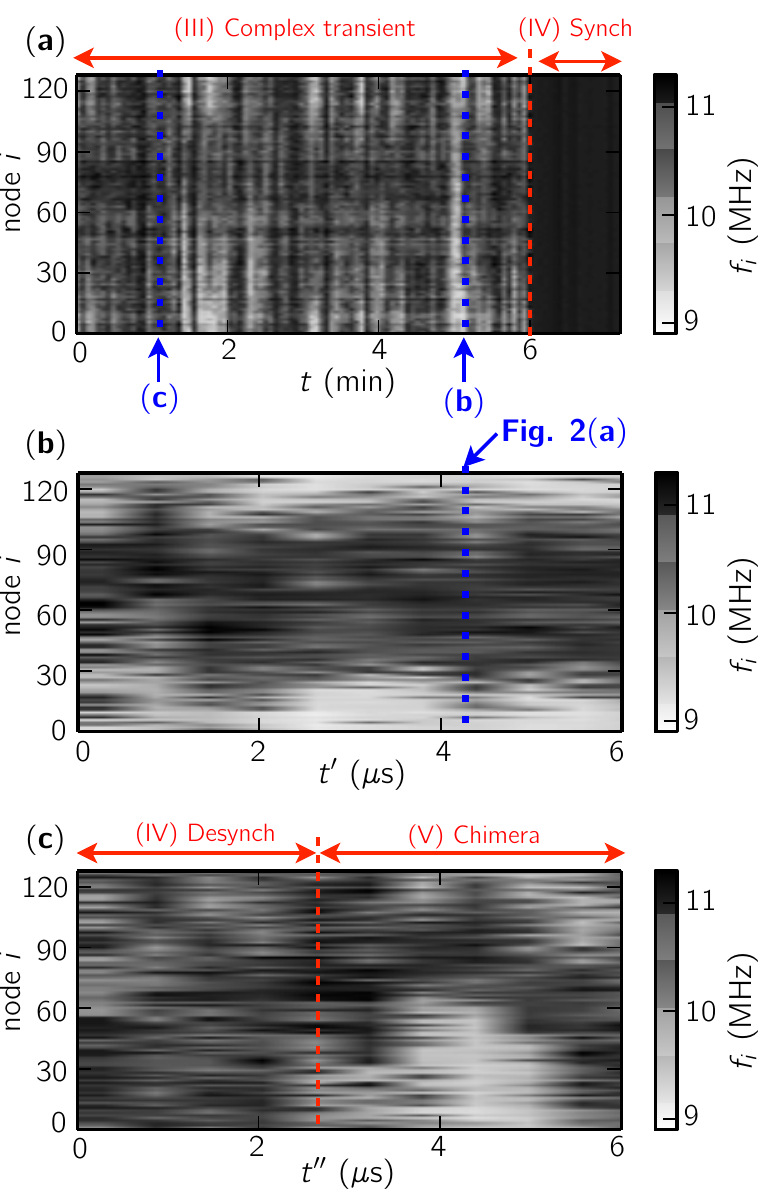}
\end{center}
\caption{\label{fig:chimera_transient}
(Color online) (a) Frequency evolution over a time period of $7\un{min}$; averaged over $6\un{\mu s}$ windows (60 oscillations) every $4\un{s}$. (b), (c) Frequency evolutions shown over a time period of $5\un{\mu s}$; averaged over $500\un{ns}$ windows (5 oscillations) with (b) $t = t' + 304\;\mathrm{s}$ and (c) $t = t'' + 56\;\mathrm{s}$.  The arrow in (b) indicates the phase measurement in Fig.~\ref{fig:exp_dynamics}(a). Parameters of the experiment as in Fig.~\ref{fig:exp_dynamics}. }
\end{figure}

The temporal evolution of the frequency is visualized in Fig.~\ref{fig:chimera_transient}(a) for a duration of ${\sim}7\un{min}$, corresponding to $\sim4\;$billion periods. For this specific realization, complex dynamics exists from time $t=0$ until $t=6\un{min}$ (marked \RNum{3}), where the frequency varies both from node to node and in time. At time $t=6\un{min}$, the dynamics collapses to a nearly synchronized state (dark gray region, marked \RNum{4}), where all but $\leq10$ oscillators have a frequency of $f=11.085\pm0.002\un{MHz}$ (compare to $f=9.14\pm0.03\un{MHz}$ for uncoupled oscillators). The remaining oscillators have a frequency different from the synchronized frequency by about $1\%$ because of heterogeneity \cite{supplement}. The time until synchronization varies considerably for different experimental runs. 

In the following, we discuss the dynamics on a microsecond timescale, at times marked in the Fig.~\ref{fig:chimera_transient}(a).

%%%%%%%%
%%%%%%%%%

Figure~\ref{fig:chimera_transient}(b) shows the frequency of the oscillators for about 60 periods after $304\un{s}$, corresponding to a millionth of the total transient. The network shows high frequencies (dark gray) for oscillator indices from $i\cong20$ to $i\cong100$ and low frequencies (light gray) for the remaining oscillators. This figure corresponds to the chimera state already identified in Fig.~\ref{fig:exp_dynamics}(a). The unsynchronized domain of the chimera state (high frequency, dark gray) moves irregularly in the network because of  finite-size effects \cite{OME10a,NKO13}; this also indicates also that the chimera state is not pinned to the network heterogeneities.

At an earlier time in the transient shown in Fig.~\ref{fig:chimera_transient}(c), the dynamics alternates between complete desynchronization and chimera states. For times $0<t<2.5\un{\mu s}$ (marked \RNum{5}), the figure shows large variations in the frequencies of neighboring nodes but with no obvious chimera domain (see phase analyses in \cite{supplement}). In the remaining time interval (marked \RNum{6}), two domains of high and low frequencies can be identified, which correspond to a chimera state that moves in the network and lasts for ${\sim}30$ oscillations. We are the first to report on this reappearance and disappearance of chimera states, which we call resurgence of chimera states.

\begin{figure}[tb]
\begin{center}
\includegraphics[width=\linewidth]{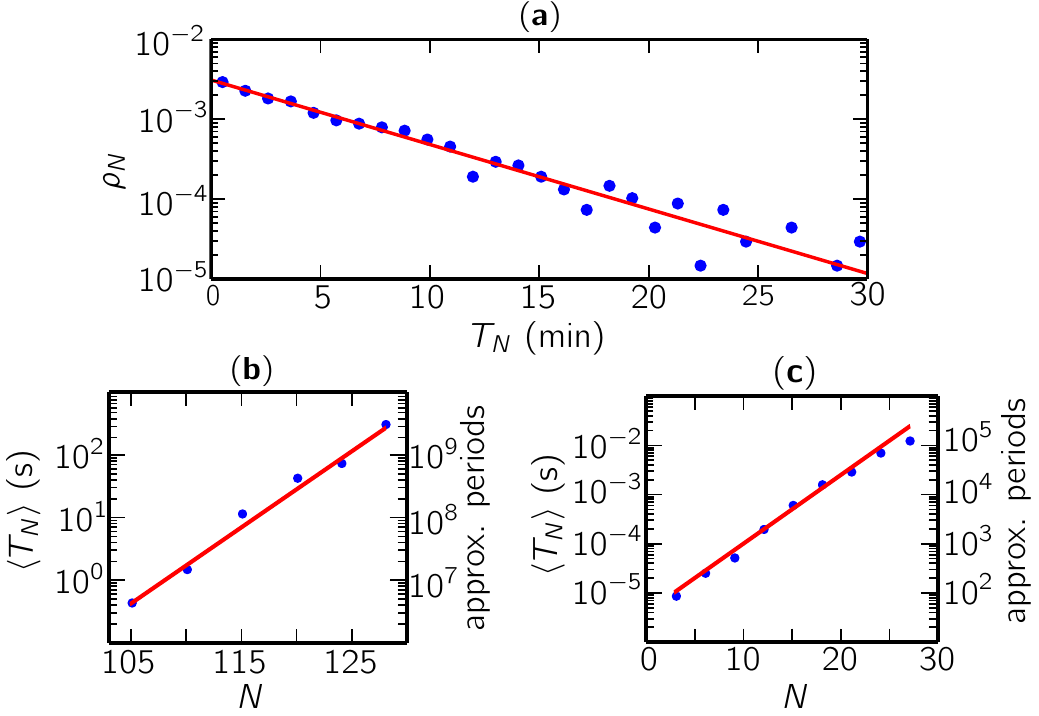}
\end{center}
\caption{\label{fig:trans_times_distr}
(Color online) (a) Histogram of transient times $T_N$ with $N=128$ from 1000 experimental acquisitions (circles) and distribution function Eq.~\eqref{eq:prob_collapse_times} (solid line). 
(b), (c)~Average transient time $\left\langle T_N\right\rangle$ as a function of $N$ measured from (b) 1000 experimental transients each, (c) 200 simulated transients each (circles). Both are fitted with Eq.~\eqref{eq:exp_scaling} (solid line) with (b) $\kappa=0.28\pm 0.10$, (c) $\kappa=0.30\pm0.08$. The right axis shows the approximate number of periods per transient. Experimental parameters $R/N\approx0.24$, $\omega_0\approx 1000/[19.7+2.9\cdot R]$ (see \cite{supplement}), $\tilde{\sigma}=\sigma\omega_0^2/\pi$ with $\sigma=0.515\pm0.018\un{ns}$, initial conditions as in Fig.~\ref{fig:exp_dynamics}. Numerical parameters are $R/N=1/3$, $\tilde{\sigma}=0.089\un{MHz}\cdot40/R$, $\alpha=0.1$ and initial conditions as in Fig.~\ref{fig:num_dynamics}. $N$ in (c) is limited by available computation time.}
\end{figure}

After a transient time $T_N$, the complex dynamics collapses to a synchronized state. We find that $T_{N}$ varies between extreme values of $T_N=1\un{s}$ and $T_N=32\un{min}$ for $N=128$ and $1{,}000$ measurements from random initializations. Different from Ref.~\cite{WIC13}, chimeras appear at every acquisition. Figure~\ref{fig:trans_times_distr}(a) shows the experimental distribution $\rho_N$ of transient times, where each dot corresponds to the normalized number of transients with a given lifetime $T_{N}$. We find that $T_{N}$ follows an exponential distribution (solid line) according to
\begin{equation}\label{eq:prob_collapse_times}
\rho_N(T_N) = \left\langle T_N\right\rangle^{-1} \exp(-T_N / \left\langle T_N\right\rangle),
\end{equation}
with the average transient time $\left\langle T_N\right\rangle=5.4\un{min}$ for $N=128$ as indicated in the figure.The exponential distribution follows analytically by considering the collapse to synchronization as a Poisson process, which occurs continuously in time at a constant average synchronization rate $\lambda=1/\langle T_N \rangle$.

Such exponential distribution has been found theoretically to describe the transient times for chimera states in the Kuramoto model under the assumption of identical oscillators~\cite{WOL11}. The appearance of the same scaling is very interesting because our experiment has heterogeneity and shows resurgence of chimeras, which are not included in previous models. 

%%%%%%%%%%
%%%%%%%%%

We measure the average transient time $\left\langle T_N\right\rangle$ for networks of different size $N$ and the same network topology. Figure~\ref{fig:trans_times_distr}(b) shows $\left\langle T_N\right\rangle$ for six different network sizes from $N=105$ to $N=128$. The average transient time $\left\langle T_N\right\rangle$ follows approximately an exponential scaling over three orders of magnitude according to
\begin{equation}\label{eq:exp_scaling}
\left\langle T_N\right\rangle \propto \exp(\kappa N),
\end{equation}
with $\kappa=0.28\pm0.10$. Using Eq.~\eqref{eq:exp_scaling} and the assumption of a Poisson process, the synchronization rate follows $\lambda \propto \exp(-\kappa N) \propto V^{-\kappa}$, which is a power law of the network state-space volume $V=(2\pi)^N$. This is plausible assuming for a single oscillator's phase-space volume $V=2\pi$ in accordance with Eq.~\eqref{eq:phase_model2}. This supertransient scaling holds for many spatially-extended systems \cite{TEL08}, neural networks~\cite{ZIL09} and networks of Kuramoto oscillators~\cite{WOL11}. This suggests that the synchronization rate $\lambda \propto V^{-\kappa}$ may be a general law for networks under certain conditions, such as nearly-identical nodes and the existence of a stable synchronized state. 

\begin{figure}[tb]
\begin{center}
\includegraphics{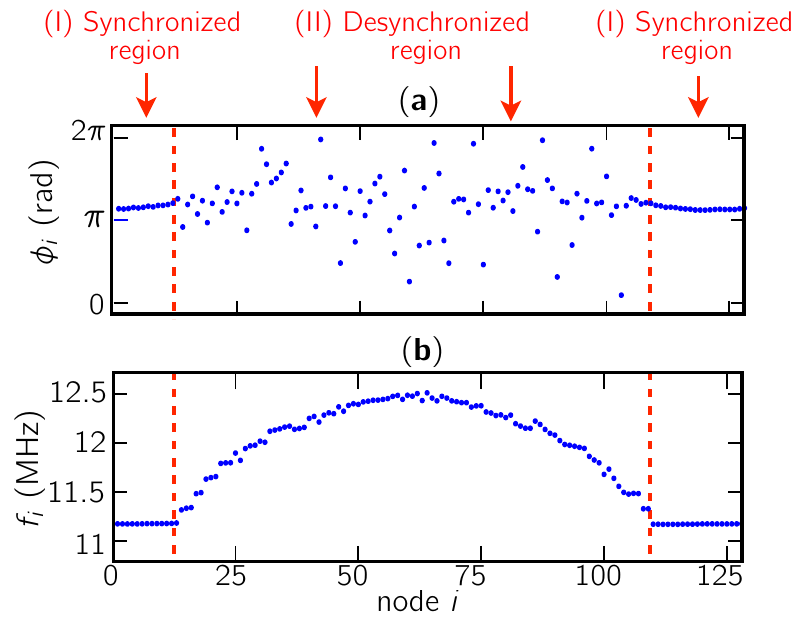}
\end{center}
\caption{\label{fig:num_dynamics}
(Color online) (a) Phases $\phi_i$ and (b) frequencies $f_i=\langle\dot{\phi_i}\rangle/(2\pi)$ of the network at $t=50\un{\mu s}$. Dynamics are obtained from numerical simulation of Eq.~\eqref{eq:phase_model2} with $N=128$, $R=42$, $\omega_0 = 2\pi\cdot9.3\un{MHz}$, $\tilde{\sigma} = 2\pi\cdot0.089\un{MHz}$, $\alpha=0.1$. Dynamics are initialized as in Ref.~\cite{ABR04} with $\phi_i = 6p\exp\left(-0.76x^2\right)$, where $p$ is a uniform random variable on $\left[-0.5,0.5\right]$ and $x=2\pi i/N-\pi$. For simplicity, we do not assume frequency heterogeneity and noise in the model. To improve simulation performance, we simulate an altered version of Eq.~\eqref{eq:phase_model2} with a continuous XOR function given by $\left\{\tanh\left[-c\sin(\phi_j)\sin(\phi_i+\alpha)\right]+1\right\}/2$ with slope $c=4$ instead of $\left|\Theta\left[\sin(\phi_j)\right] -\Theta\left[\sin(\phi_i+\alpha)\right]\right|$.
}
\end{figure}

We study the network dynamics numerically using the simplified model in Eq.~\eqref{eq:phase_model2}. Analogous to Fig.~\ref{fig:exp_dynamics}, Fig.~\ref{fig:num_dynamics} shows the dynamics in phase and frequency representations. We use a different coupling range of $R=42$ ($R/N\approx 1/3$) than in the experiment because the value used in the experiment ($R=30$) does not lead to chimera states in the simulation. The figure shows a chimera state with co-existence of a synchronized and desynchronized domains (see also the explanation for Fig.~\ref{fig:exp_dynamics}). The model also reproduces the characteristic scaling of the transient of Eq.~\eqref{eq:exp_scaling}, as shown in Fig.~\ref{fig:trans_times_distr}(c) with $\kappa=0.30\pm0.08$, which is similar to Fig.~\ref{fig:trans_times_distr}(b). Both results suggest that the model is well suited to describe our experiment qualitatively. 

%%%%%%%%
%%%%%%%%%

The model is, however, only a first step towards a complete theoretical description of the experimental dynamics because of several differences. First, the simulation shows chimera states for the entire transient and does not show the resurgence of chimeras as in the experiment \cite{supplement}. Second, the simulation (experiment) collapses to a synchronized (nearly synchronized) state, where nodes are phase and frequency synchronized (nearly frequency synchronized but not phase synchronized) after the transient \cite{supplement}. Third, chimera states appear in different parameter regions in the model and experiment. 

These differences may be caused by heterogeneity in the experiment $\alpha_{ij}\neq \mathrm{const}$, while $\alpha_{ij}=\mathrm{const}$ is assumed in the model. Specifically, the experiment implements heterogeneous wiring leading to differences in link delays \cite{supplement}. Further more, differences may be caused by noise and frequency heterogeneity of $0.3\%$, and transmission delays along the links ($<5\un{ns}$) in the experiment. Future work has to fill this gap to uncover the underlying mechanism.

In conclusion, we study a network of Boolean phase oscillators that approximately follows equations similar to the Kuramoto model \cite{KUR02a}. Large experimental networks of up to 128 non-locally coupled Boolean phase oscillators show complex transient dynamics, where chimera states disappear and reappear called resurgence of chimera states, which is not yet theoretically understood. The dynamics collapses to a synchronized state after a long transient, which can be modeled by a Poisson process with an average lifetime scaling exponentially with the network size, as  predicted theoretically in coupled Kuramoto oscillators \cite{WOL11}. The appearance of supertransient scaling in our experimental networks provides further evidence that this scaling could be a general feature of certain networks. 
Our work motivates future experimental studies, such as transient scaling in spiking neural networks and control of chimera states \cite{SIE14c}. 

\section{Acknowledgments}
The authors gratefully acknowledge financial support by the U.S. Army Research Office within Grant W911NF-12-1-0099. D.P.R. and E.S. acknowledge the DFG for financial support in the framework of SFB910. D.R acknowledges Fondation Sup{\'e}lec for financial support. D.P.R. thanks Philipp H\"ovel for helpful discussions.

%\bibliographystyle{apsrev4-1}
%\bibliography{references}
%\bibliography{ref}
%merlin.mbs apsrev4-1.bst 2010-07-25 4.21a (PWD, AO, DPC) hacked
%Control: key (0)
%Control: author (8) initials jnrlst
%Control: editor formatted (1) identically to author
%Control: production of article title (-1) disabled
%Control: page (0) single
%Control: year (1) truncated
%Control: production of eprint (0) enabled
%

\end{document}